\documentclass[secnumarabic, graphics,floatfix, nofootinbib,tightenlines,nobibnotes, aps, prl, 12pt]{revtex4-2}
\usepackage[english]{babel}
\usepackage[utf8]{inputenc}
\usepackage{amsfonts}
\usepackage{multirow}
\usepackage{graphicx}
\usepackage{epstopdf}
\usepackage{hyperref}
\usepackage{latexsym}
\usepackage{amsmath}
\usepackage{amssymb}

\usepackage[utf8]{inputenc}
\hypersetup{
    colorlinks=true,
    linkcolor=red,
    citecolor=blue,
}
\usepackage{color}
\usepackage[T1]{fontenc}
\usepackage{txfonts}
\usepackage{mathrsfs}
\usepackage[center]{subfigure}

\begin{document}
 \setcounter{secnumdepth}{2}
 \newcommand{\bq}{\begin{equation}}
 \newcommand{\eq}{\end{equation}}
 \newcommand{\bqn}{\begin{eqnarray}}
 \newcommand{\eqn}{\end{eqnarray}}
 \newcommand{\nb}{\nonumber}
 \newcommand{\lb}{\label}
 
\title{An implementation of the matrix method using Chebyshev grid}

\author{Shui-Fa Shen$^{1, 2, 3}$}\email[E-mail: ]{shuifa.shen@inest.cas.cn}
\author{Wei-Liang Qian$^{4, 5, 6}$}\email[E-mail: ]{wlqian@usp.br (corresponding author)}
\author{Hong Guo$^{6, 7}$}\email[E-mail: ]{gh710105@gmail.com }
\author{Shao-Jun Zhang$^{8}$}\email[E-mail: ]{sjzhang@zjut.edu.cn}
\author{Jin Li$^{9}$}\email[E-mail: ]{cqstarv@hotmail.com}

\affiliation{$^{1}$ School of intelligent manufacturing, Zhejiang Guangsha Vocational and Technical University of Construction, 322100, Jinhua, Zhejiang, China}
\affiliation{$^{2}$ School of Electronic, Electrical Engineering and Physics, Fujian University of Technology, 350118, Fuzhou, Fujian, China}
\affiliation{$^{3}$ Hefei Institutes of Physical Science, Chinese Academy of Sciences, 230031, Hefei, Anhui, China}
\affiliation{$^{4}$ Escola de Engenharia de Lorena, Universidade de S\~ao Paulo, 12602-810, Lorena, SP, Brazil}
\affiliation{$^{5}$ Faculdade de Engenharia de Guaratinguet\'a, Universidade Estadual Paulista, 12516-410, Guaratinguet\'a, SP, Brazil}
\affiliation{$^{6}$ Center for Gravitation and Cosmology, School of Physical Science and Technology, Yangzhou University, 225002, Yangzhou, Jiangsu, China}
\affiliation{$^{7}$ School of Aeronautics and Astronautics, Shanghai Jiao Tong University, 200240, Shanghai, China}
\affiliation{$^{8}$ Institute for Theoretical Physics \& Cosmology, Zhejiang University of Technology, 310032, Hangzhou, Zhejiang, China}
\affiliation{$^{9}$ College of Physics, Chongqing University, 401331, Chongqing, China}

\date{Oct. 30th, 2022}

\begin{abstract}
In this work, we explore the properties of the matrix method for black hole quasinormal modes on the nonuniform grid.
In particular, the method is implemented to be adapted to the Chebyshev grid, aimed at effectively suppressing Runge's phenomenon.
It is found that while such an implementation is favorable from a mathematical point of view, in practice, the increase in precision does not necessarily compensate for the penalty in computational time.
On the other hand, the original matrix method, though subject to Runge's phenomenon, is shown to be reasonably robust and suffices for most applications with a moderate grid number.
In terms of computational time and obtained significant figures, we carried out an analysis regarding the trade-off between the two aspects.
The implications of the present study are also addressed.

\end{abstract}

\maketitle

\newpage

\section{Introduction}\label{section1}

Recently, Jaramillo {\it et al.}~\cite{agr-qnm-instability-07, agr-qnm-instability-13, agr-qnm-instability-14} demonstrated the possibility of generic instability of black hole quasinormal modes (QNM) against high-frequency perturbations. 
Based on the studies initiated by Nollet, Price, and others~\cite{agr-qnm-35, agr-qnm-36, agr-qnm-50, agr-qnm-lq-03}, the work was carried out from a rather general perspective, using primarily the pseudospectrum method~\cite{book-pseudospetrum-method-Sjostrand}. 
Under randomized high-frequency perturbations, the onset of a universal instability is indicated chiefly by the high-overtone modes, migrating toward the real axis.
The pseudospectrum analysis has been explored for the charged~\cite{agr-qnm-instability-14}, rotating black holes~\cite{agr-qnm-instability-07}, and horizonless compact objects~\cite{agr-qnm-instability-25}.
Recently, it was further pointed out~\cite{agr-qnm-instability-15, agr-qnm-instability-16} that even the fundamental mode can be destabilized under insignificant perturbations.
Nonetheless, one crucial feature of such analysis is the notion of {\it operator norm}, which resides on a specific and physically relevant definition of scalar product~\cite{agr-qnm-instability-06}.
Recently, the latter aspect was scrutinized further by Yang and Zhang~\cite{agr-qnm-instability-26}.

In terms of physical implication, the most substantial impact of the spectral instability is related to gravitational-wave spectroscopy, particularly given the prospects regarding ongoing space-borne interferometric antennae.
Indeed, in the literature, much effort has been devoted aiming at extracting information from the empirical waveforms, known as the black hole spectroscopy~\cite{agr-BH-spectroscopy-05, agr-BH-spectroscopy-06, agr-BH-spectroscopy-10, agr-BH-spectroscopy-18, agr-BH-spectroscopy-20, agr-BH-spectroscopy-36}.
In this regard, as pointed out in~\cite{agr-qnm-instability-13}, such spectral instability might lead to immediate implications in gravitational-wave astronomy.
This is because a black hole or other compact object is always merged in a realistic environment, and therefore the resulting QNM spectrum might deviate significantly from their counterparts, namely, the pure mathematical solutions in the vacuum.
Subsequently, one might expect that the emanated gravitational waves, especially at the late stage, deviate from the theoretical estimations and undermine the analysis's precision.
Intriguingly, the robustness of the black hole spectroscopy does not seem to be undermined from the time-domain perspective~\cite{agr-qnm-instability-16}.
In particular, the initial pulses of the time-domain QNM waveforms agree with the unperturbed fundamental mode.
If the fit is carried out for the entire time domain waveforms, the modified fundamental mode can be largely retrieved, but the process cannot be extended to extract higher overtones.
On the other hand, by employing the polynomial initial data, the first few higher overtone was successfully retrived~\cite{agr-qnm-instability-13}. 
Nonetheless, it was recently speculated that spectral instability might have a limited effect in the scattered gravitational waves~\cite{agr-qnm-instability-18}.

Besides black hole spectroscopy, the underlying physics associated with the spectral instability of quasinormal modes against high-frequency perturbations is a relevant phenomenon regarding the basic properties of black holes.
Among others, it might be of interest for the gauge-gravity duality.
From a practical point of view, the above studies call for a numerical algorithm for black hole QNMs with unprecedented precision.
The relevant approaches consist of two types.
The first one is to evaluate the quasinormal frequencies semi-analytically or numerically directly.
In the literature, the shooting method~\cite{agr-qnm-16}, continued fraction method~\cite{agr-qnm-continued-fraction-01, agr-qnm-star-08} have been extensively employed to obtain the QNMs at the desirable precision.
The second approach is to extract the QNM from the time-domain profile obtained by the finite difference method~\cite{agr-qnm-finite-difference-01}.
The extraction is performed using the Prony method~\cite{agr-qnm-55}.
The explicit calculations for a specific form of potential perturbation are relevant since the pseudospectrum method can only estimate the bound of the instability.
In this regard, it is meaningful to explore further the effect of the relevant perturbations on the non-selfadjoint operator in the context of black hole physics.
Among others, effective potentials with discontinuity might play an interesting role as a mathematically simple and physically relevant model.

Proposed by some of us, the matrix method~\cite{agr-qnm-lq-matrix-01,agr-qnm-lq-matrix-02,agr-qnm-lq-matrix-03,agr-qnm-lq-matrix-04,agr-qnm-lq-matrix-05,agr-qnm-lq-matrix-06, agr-qnm-lq-matrix-08} is an approach that turns the problem of solving the master equation for the quasinormal frequencies into a non-standard matrix eigenvalue problem.
Like the well-known continued fraction method~\cite{agr-qnm-continued-fraction-01}, Taylor expansion is utilized to rewrite the wave function.
The main difference, however, is that the matrix method makes use of a series of grid points instead of a given position~\cite{agr-qnm-lq-matrix-01}.
The method is shown to be capable of handling the metrics with spherical symmetry~\cite{agr-qnm-lq-matrix-02}, axial symmetry~\cite{agr-qnm-lq-matrix-03}, as well as the system composed of various coupled degrees of freedom~\cite{agr-qnm-lq-matrix-07}.
It can be adapted to different boundary conditions~\cite{agr-qnm-lq-matrix-04} and dynamic black hole spacetimes~\cite{agr-qnm-lq-matrix-05}.
More recently, the method was generalized~\cite{agr-qnm-lq-matrix-06} to the potentials containing discontinuity and aimed to the higher orders~\cite{agr-qnm-lq-matrix-08}.
The approach provides reasonable accuracy and efficiency and has been utilized in various studies~\cite{Destounis:2018utr, Destounis:2018qnb, Panotopoulos:2019tyg, Destounis:2019zgi, Hu:2019cml, Cardoso:2017soq, Liu:2019lon, agr-qnm-50, Shao:2020gwr, Lei:2021kqv, Zhang:2022roh, Li:2022khq, Shao:2022oqv, Mascher:2022pku}.

It is well-known that polynomial interpolation based on a uniform grid is subject to Runge's phenomenon.
The latter is typically featured by significant oscillations at the edges of the interpolation interval, attributed to an uncontrolled increase of the Lebesgue constant~\cite{book-approximation-theory-Rivlin}.
The appearance of Runge's phenomenon in the matrix method~\cite{agr-qnm-lq-matrix-10} is due mainly to the choice of the uniform grid but not the method itself.
In other words, it is interesting to generalize the approach to adopt some well-known nonuniform grids, such as the Chebyshev one.
The present study is primarily motivated by the above considerations.
The matrix method is updated to adopt the Chebyshev grid. 
The new version of the code is then utilized to evaluate the black hole QNMs, and the results are compared to the previous versions of the method.
Moreover, investigate the precision and efficiency of the novel implementation.

The remainder of the paper is organized as follows.
In the following section, we briefly review the matrix method.
We discuss the implementation of the Chebyshev grid in Sec.~\ref{section3}.
In Sec.~\ref{section4}, we analyze the new version of the code by evaluating some well-known black hole QNM problems.
By taking the axial gravitational and scalar perturbations in asymptotically flat and anti-de Sitter spacetimes as examples, we elaborate on the precision and efficiency of the code.
It is pointed out that even though the new version provides improved precision and is free of Runge's phenomenon, the efficiency of the code is somewhat hampered by the nonuniform grid.
The concluding remarks are given in Sec.~\ref{section5}.

\section{The matrix method}\label{section2}

As discussed above, one of the main features of the matrix method~\cite{agr-qnm-lq-matrix-01,agr-qnm-lq-matrix-02,agr-qnm-lq-matrix-03}, distinct from other approaches, resides in the procedure where a series of Taylor expansions is carries out for the wavefunction on a grid.
Although the uniform grid was utilized in most implementations, it is not necessarily an obligation.
When it applies, this feature guarantees flexibility, which is essential in order to achieve a higher degree of precision.
By adopting a grid, we discretize the master equation of the perturbations.
Subsequently, it is rewritten as a non-standard eigenvalue equation of the complex frequency in the form of a matrix equation.
The latter can be readily solved by using some standard algorithm.

For the black hole QNMs, the master equation is typically of the following Schr\"odinger-type~\cite{agr-qnm-review-01, agr-qnm-review-02, agr-qnm-review-03, agr-qnm-review-06}:
\bqn
\lb{qnmeq}
\frac{d^2}{dr_*^2}\Phi(r)+\left[\omega^2-V(r)\right]\Phi(r)=0 ~,
\eqn
where $r_*=\int{dr/F(r)}$ is the tortoise coordinate, $F(r)$ is governed by the specific metric, $\omega$ and $V(r)$ are the quasinormal frequency and effective potential respectively.
Although the above master equation generally applies to a wide variety of metrics, it is noted that for the case of either rotating black holes or metrics expressed in terms of Eddington-Finkelstein coordinates, appropriate modification has to be made to the formulation.
The specific forms of the master equation and the corresponding boundary conditions in the latter cases were further elaborated in~\cite{agr-qnm-lq-matrix-04}.

In order to discretize the differential equation, Eq.~\eqref{qnmeq}, an essential step is to formally discretize an arbitrary wave function defined on an interval $[x_\mathrm{min}, x_\mathrm{max}]$.
To proceed, one introduce a grid of $N$ points: ${\cal X}\equiv \left(x_1, x_2, \cdots, x_N\right)$, where $x_1=x_\mathrm{min}$ and $x_N=x_\mathrm{max}$.
Subsequently, Taylor expansion is carried out at individual nodes for the unknown waveform up to $N$th order.
Without loss of generality, let us expand the waveform about $x_2$ to $x_1,x_3,x_4,\cdots,x_N$, and write down the obtained $N-1$ linear relations between function values and their derivatives in the following matrix form
\bqn
\lb{2}
\Delta{\cal F}=M {\cal D} ,
\eqn
where
 \bqn
 \lb{3}
\Delta{\cal F}=\left(
    f(x_1)-f(x_2),
    f(x_3)-f(x_2),
    \cdots,
    f(x_j)-f(x_2),
    \cdots,
    f(x_{N})-f(x_2)
\right)^T ,
\eqn

\bqn
\lb{4}
M= \left(
  \begin{array}{cccccc}
    x_1-x_2 & \frac{(x_1-x_2)^2}{2} &\cdots & \frac{(x_1-x_2)^i}{i!} &\cdots & \frac{(x_1-x_2)^{N-1}}{{(N-1)}!} \\
    x_3-x_2 & \frac{(x_3-x_2)^2}{2} &\cdots & \frac{(x_3-x_2)^i}{i!} &\cdots & \frac{(x_3-x_2)^{N-1}}{{(N-1)}!} \\
        \cdots & \cdots & \cdots & \cdots & \cdots &\cdots \\
    x_j-x_2 & \frac{(x_j-x_2)^2}{2} &\cdots & \frac{(x_j-x_2)^i}{i!} &\cdots & \frac{(x_j-x_2)^{N-1}}{{(N-1)}!} \\
        \cdots & \cdots & \cdots & \cdots & \cdots &\cdots \\
    x_{N}-x_2 & \frac{(x_{N}-x_2)^2}{2} &\cdots & \frac{(x_{N}-x_2)^i}{i!} &\cdots & \frac{(x_{N}-x_2)^{N-1}}{{(N-1)}!} \\
  \end{array}
\right) ,
\eqn

\bqn
\lb{5}
{\cal D}= \left(
    f'(x_2),
    f''(x_2),
    \cdots,
    f^{(k)}(x_2),
    \cdots,
    f^{({N})}(x_2)
\right)^T .
\eqn

The discretization of the master equation can be achieved by solving Eq.~\eqref{2} for the components of $\cal D$.
Specifically, all the derivatives at $x=x_2$ can be expressed in terms of the function values.
Using Cramer's rule, we have
\bqn
\lb{5a}
f'(x_2)= \det(M_1)/\det(M),\nb\\
f''(x_2)= \det(M_2)/\det(M), \eqn
where $M_i$ is a matrix obtained by replacing the $i$th column of $M$ by the column vector $\Delta{\cal F}$.
Now, by permuting the $N$ grid points, $x_1, x_2, \cdots, x_N$, one obtains all the derivatives at the nodes in terms of the waveform evaluated at those points.
By substituting these matrix expressions into the master equation Eq.~\eqref{qnmeq}, one derives its discretized version.

In practice, two issues remain to be adequately accounted for: the boundary conditions and radial coordinate variable transform.
First, the physically relevant boundary conditions have to be implemented.
For the black hole QNMs, the waveform must be asymptotically in-going at the horizon and out-going at spatial infinity.
Similar to the continued fraction method, this is carried out by factoring out the asymptotical waveforms from the function $\Phi(r_*)$ so that the resultant function $\Psi(r_*)$ behaves moderately at the boundaries.
Second, we change the domain of the radial variable to a finite range by introducing $x=x(r_*)$, and thus $\Psi(r_*)\to F(x)$.
For convenience, we choose the domain to be $x\in[0,1]$ so that the boundaries correspond to the points $x=0$ and $x=1$, respectively.
In the original implementation of the method~\cite{agr-qnm-lq-matrix-02, agr-qnm-lq-matrix-03}, it was suggested that the equations which are closer to the boundaries are less precise, and an additional transform $F(x)=x(1-x)R(x)$ are introduced.
For instance, in the case of the asymptotically flat Schwarzschild spacetime below, we choose to replace the first and the last line in the matrix equation by equating them to zero.
Recently, it was pointed out~\cite{agr-qnm-lq-matrix-08, agr-qnm-lq-matrix-10} that such a procedure was not entirely necessary.

By putting the above pieces together, the master equation Eq.~\eqref{qnmeq} is rewritten into
\bqn
\lb{matrixEq}
G \ {\cal F} = 0 ,
\eqn
where the column matrix 
\bqn
{\cal F}=(f_1,f_2,\cdots,f_i,\cdots,f_N)^T ,
\eqn
and $G$ is an $N\times N$ matrix.
It is the discretized version of a linear operator determined by Eq.~\eqref{qnmeq}, which depends on the quasinormal frequency $\omega$ and $x$.
Therefore, we have descretized the master equation by introducing $N$ interpolation points $x_i$ with $i=1, 2, \cdots, N$ for the interval $x\in[0,1]$.
The matrix equation Eq.~\eqref{matrixEq} indicates that ${\cal F}$ is the eigenvector of $G$, which implies
\bqn
\lb{qnmDet}
\det G(\omega)=0~.
\eqn

Instead of Eq.~\eqref{5a}, one may also evaluate the differentiation matrices using the lagrangian polynomials.
Numerically, it turns out that Eq.~\eqref{5a} is more efficient for higher-order derivatives, as its computational time does not increase with the order.
Nonetheless, as the differentiation matrices only need to be evaluated once, it does not pose an issue in practice. 
The following section will generalize the matrix method to adopt the Chebyshev grid.

\section{Implementation of Chebyshev grid} \lb{section3}

When a uniform grid is utilized, it is well-known that undesirable oscillations emerge at the edge of the interpolation interval, owing to Runge's phenomenon~\cite{book-approximation-theory-Rivlin}.
The bound of the amount of deviation can be estimated by the Lebesgue constant associated with the interpolation process on the grid.  
Effectively, the Lebesgue constant provides a quantitative measure of the interpolant of a function in comparison with the best polynomial approximation at a given degree.
The quantities can be evaluated using the Weierstrass approximation theorem to obtain the remainder in the Lagrange interpolant.
The resultant upper bound is governed by two factors~\cite{book-numerical-analysis-Burden-Faires}.
First, the specific form of the nodal function is governed by the underlying grid.
Second, the magnitude of the derivative of the waveform.
In terms of a uniform grid, according to Turetski~\cite{zmath-Runge-Chebyshev-02}, the Lebesgue constant was found to increase exponentially with the number of the grid, causing Runge's phenomenon.

In the literature, a well-established recipe for suppressing Runge's phenomenon is the Chebyshev grid~\cite{book-approximation-theory-Rivlin, book-numerical-analysis-Burden-Faires}.
The latter accomplishes its goal by minimizing the Lagrange interpolation error.
It attains the minimum of the maximum of the nodal functions assigned to an arbitrary grid.
There is also an elegant physical interpretation for the use of the Chebyshev grid, and it leads to a more restrictive Runge region, defined as the interior of an {\it equalpotential} curve~\cite{book-approximation-theory-Trefethen-01}.  
In principle, exponential convergence can be achieved.
The Chebyshev grid is defined on the interval $[-1, 1]$ as follows
\bqn\label{ChebGrid}
y_j^{\mathrm{Cheb}} =\cos\left(\frac{\pi j}{N}\right) , \ \ j=0, 1, \cdots N .
\eqn 
The above nonuniform grid can be viewed as the projections onto the interval $[-1, 1]$ from equally spaced points along a unit circle.
They are the extreme points of the Chebyshev polynomials, also referred to in the literature as the Chebyshev points of the second kind.
The main feature of such a distribution is that the nodes mostly cluster at the two edges of the interpolation interval by a density $\propto N^2$.

Implementing the Chebyshev grid in the matrix method is straightforward.
One carried out the following transformation from $x$ to $y$:
\bqn\label{ChebGrid2x}
y = 2x -1 ,
\eqn 
and rewrites the master equation Eq.~\eqref{matrixEq} accordingly.
The quasinormal frequencies are obtained by solving the modified version of Eq.~\eqref{qnmDet}.
As an example, for $N=5$, the first derivative of the waveform  
\bqn\label{calFp}
{\cal F}' = ({f_1}',{f_2}',\cdots,{f_i}',\cdots,{f_N}')^T 
\eqn
is obtained by multiplying the following coefficient matrix, from Eq.~\eqref{5a}, to the waveform at the grid, which reads
\bqn\label{d1Matrix5}
{\cal F}' = 
\begin{pmatrix}
-\frac{11}{2}&-\frac{2}{-1+\frac{1}{\sqrt{2}}}&-2&\frac{2}{1+\frac{1}{\sqrt{2}}}&-\frac12\\
-\frac{1}{2\left(1-\frac{1}{\sqrt{2}}\right)}&\frac{1}{\sqrt{2}}&\sqrt{2}&-\frac{1}{\sqrt{2}}&\frac{1}{2\left(1+\frac{1}{\sqrt{2}}\right)}\\
\frac12&-\sqrt{2}&0&\sqrt{2}&-\frac12\\
-\frac{1}{2\left(1+\frac{1}{\sqrt{2}}\right)}&\frac{1}{\sqrt{2}}&-\sqrt{2}&-\frac{1}{\sqrt{2}}&-\frac{1}{2\left(-1+\frac{1}{\sqrt{2}}\right)}\\
\frac12&-\frac{2}{1+\frac{1}{\sqrt{2}}}&2&-\frac{2}{1-\frac{1}{\sqrt{2}}}&\frac{11}{2}
\end{pmatrix}
\ {\cal F} . 
\eqn 
The remainder of the procedure is carried out mainly following what has been discussed in the previous section.
In the next section, numerical results will be presented.

Nonetheless, the specific implementation involves some subtlety.
The elements of the differentiation matrices are generally not rational numbers, and therefore, the solution of the eigenvalue equation becomes increasingly time-consuming.
In this regard, we adopt the following two different implementations.
The first one, referred to as MV1, is to execute precisely the described algorithm.
Second, we also propose an alternative choice as follows.
Instead of evaluating the matrices such as Eq.~\eqref{d1Matrix5} exactly on the Chebyshev grid, one chooses the neighboring grid points whose coordinates are rational numbers.
To be specific, for $N=5$, we will use the nodes
\bqn
{\cal X}' = \left(-1, \frac{29}{41}, 0, \frac{29}{41}, 1\right) ,
\eqn
which is consistent with the corresponding Chebyshev grid
\bqn
{\cal X}_{\mathrm{Cheb}} = \left(-1, -\frac{1}{\sqrt{2}}, 0, \frac{1}{\sqrt{2}}, 1\right) ,
\eqn
by three significant figures, in conjunction with the corresponding differentiation matrices to replace those given by Eq.~\eqref{d1Matrix5}.
This alternative approach is essentially inspired by the spirit of the {\it mock-Chebyshev} grid~\cite{zmath-Runge-mock-Chebyshev-01, zmath-Runge-mock-Chebyshev-02}, mathematically based on the Rakhmanov theorem~\cite{zmath-Runge-Chebyshev-05}.
Such a method proposes to perform a polynomial interpolation of degree $P$ using a given number of $N = P$ nodes.
Instead of using the Chebyshev grid, the specific choice of the grid points are those from a uniform distribution but mimicking the Chebyshev grid as much as possible.
As pointed out above, the nodes of the Chebyshev grid are clustering near the end of the interpolation interval.
This feature is understood to be the essence of effectively suppressing the potential oscillations of the interpolant, according to Rakhmanov's theorem.
In particular, the density of the grid points is quadratic in $N$.
In fact, it was shown~\cite{zmath-Runge-Chebyshev-05} that convergence could be achieved at the same rate as the Chebyshev grid.
In our case, the grid in question is featured by a nonuniform distribution.
However, motivated by the above considerations, the grid points are chosen to be as close as the Chebyshev one.
By attaining rational numbers, one expects that the algorithm will become less expensive in computational time.
The second approach will be referred to as MV2 below.

\section{Numerical results}\lb{section4}

For the numerical calculations, we make use of a personal computer configured with an {\it Intel} {\it Core} i7-11700 CPU @2.50GHz and $64.0$ GB {\it Kingstone} 2400 DDR4 memory. 
The numerical algorithms are implemented using {\it Mathematica} 13.0 on {\it Windows} 11 pro 22H2, under which the computational time is estimated. 
In the remainder of this paper, calculations are carried out for the fundamental mode of two specific cases.
We considered the axial gravitational perturbations in asymptotically flat spacetime for the first case.
The second case involves the scalar perturbations in asymptotically anti-de Sitter spacetime.
The conclusion drawn in this section has also been validated for other types of perturbations with different angular momenta.

We consider the QNMs in Regge-Wheeler potential for $3+1$ Schwarzschild black holes, which reads~\cite{agr-qnm-review-03}
\begin{equation}\label{eqRW}
V_{\mathrm{eff}}=V_{\mathrm{RW}}(r)\equiv F(r)\left[\frac{\ell(\ell+1)}{r^2}+(1-s^2)\left(\frac{r_0}{r^3}+\frac{4-s^2}{2L^2}\right)\right] ,
\end{equation}
where
\bqn
F(r) = 1-\frac{r_0}{r}+\frac{r^2}{L^2}
\eqn
where $r$ is the radial coordinate, related to the tortoise coordinate by $r_*=\int\frac{dr}{1-\frac{r_h}{r}}$, $L$ represents the curvature radius of the anti-de Sitter spacetime, $r_0$ is the location of the horizon for asymptotically flat spacetimes, and $\ell$ corresponds to the angular momentum.

The axial gravitational perturbations in asymptotically flat space are carried out with the parameters $s=-2$, $r_0=r_h=1$, $\ell = 2$, and $L\to \infty$. 
The master equation Eq.~\eqref{qnmeq}, therefore can be written as
\bqn\label{eqRWr}
\begin{aligned}
&\left[r^2(r-r_h)\right]\Phi''(r)+\left[r(r_h+2ir^2\omega -4ir_h^2\omega)\right]\Phi'(r)\\
&+\left[-\ell(\ell+1) r+r_h(-1+s^2+4r_h^2\omega^2+4r_h\omega(i+r\omega))\right]\Phi(r)=0 .
\end{aligned}
\eqn

We introduce the variable $x$
\bqn
x = \frac{r-r_h}{r} ,
\eqn
which takes ``$0$'' and ``$1$'' at the horizon and spatial infinity, and
\bqn
y = 2\frac{r-r_h}{r} - 1 ,
\eqn
which manifestly attains ``$\mp 1$'' at the relevant boundaries.
The asymptotical behavior of the wavefunction is subtracted by defining
\bqn
\Phi(r) = (r-r_h)^{-i\omega r_h}e^{i\omega(r-r_h)r^{2i\omega r_h}}\Psi(r) .
\eqn
As a result, the master equation Eq.~\eqref{eqRWr} gives
\bqn\label{eqRWx}
\begin{aligned}
&\left[(x-1)^2 x\right] R''(x)+\left[1-2 i r_h\omega+x^2(3-4ir_h\omega)+x(-4+8ir_h\omega)\right]R'(x)\\
&+\left[-1-\ell(\ell+1)-s^2(x-1)+x-4ir_h\omega(x-1)+8r_h^2\omega^2-4r_h^2x\omega^2\right]R(x)=0 
\end{aligned}
\eqn
in $x$, where $R(x)=\Psi(r(x))$ for the standard matrix method and 
\bqn\label{eqRWy}
\begin{aligned}
&\left[(y-1)^2(y+1)\right]P''(y)+\left[-1+4ir_h\omega+y^2(3-4ir_h\omega)+y(-2+8ir_h\omega)\right]P'(y)\\
&+\left[-1-2\ell(\ell+1)-s^2(y-1)+y-4ir_h\omega(y-1)+12r_h^2\omega^2-4r_h^2y\omega^2\right]P(y)=0 .
\end{aligned}
\eqn
in variable $y$ for Chebyshev grid, where $P(y)=\Psi(r(y))$.

As a reference for the accurate value regarding the calculations presented below, the quasinormal frequency for the axial gravitational QNMs in asymptotically flat spacetime is also evaluated by employing the continued fraction method~\cite{agr-qnm-continued-fraction-01} at $300$th order.
The value is found to be 
\bq
\omega_\mathrm{CF} = 0.74734336883598689863 - 0.17792463137781263197 i .\nb
\eq

\begin{table*}[ht]
\caption{\label{tab1} The calculated fundamental mode employing the matrix method on the Chebyshev grid (MV1) for the axial gravitational perturbations in asymptotically flat spacetime by using different grid numbers.}
\centering
\begin{tabular}{c c c}
         \hline\hline
\text{Grid number }$N$ & \text{Time (seconds)} & $\omega$  \\
\hline
\hline
5  & 0.032 &~~~$0.74696102522880045213-0.18372752541665119796 i$~~~\\
10 & 0.21 &~~~$0.74737336975789860418-0.17790765219597412314 i$~~~\\
15 & 0.86 &~~~$0.74734325056951374797-0.17792402499141445329 i$~~~\\
20 & 2.19 &~~~$0.74734335602295398730-0.17792461762436544278 i$~~~\\
25 & 3.69 &~~~$0.74734336827014213102-0.17792463082022346651 i$~~~\\
\hline
\hline
\end{tabular}
\end{table*}

\begin{table*}[ht]
\caption{\label{tab2} The same as Tab.~\ref{tab1} but employing the matrix method implemented on the {\it rational} Chebyshev grid (MV2) as described in the text.}
\centering
\begin{tabular}{c c c}
         \hline\hline
\text{Grid number }$N$ & \text{Time (seconds)} & $\omega$  \\
\hline
\hline
5  & 0.021 &~~~$0.74696453393389959800-0.18373270485576667588 i$~~~\\
10 & 0.053 &~~~$0.74737354746080685985-0.17790716923153434583 i$~~~\\
15 & 0.12 &~~~$0.74734324283660233818-0.17792403099535900868 i$~~~\\
20 & 0.28 &~~~$0.74734335560166591054-0.17792461760411252349 i$~~~\\
25 & 0.43 &~~~$0.74734336827173259938-0.17792463082421838607 i$~~~\\
\hline
\hline
\end{tabular}
\end{table*}

\begin{table*}[ht]
\caption{\label{tab3} The calculated fundamental mode employing the original matrix method on the uniform grid (MVO) for the axial gravitational perturbations in asymptotically flat spacetime by using different grid numbers.}
\centering
\begin{tabular}{c c c}
         \hline\hline
\text{Grid number }$N$ & \text{Time (seconds)} & $\omega$  \\
\hline
\hline
5  & 0.0099 &~~~$0.74467725692444945148-0.17949985765049388355 i$~~~\\
10 & 0.023 &~~~$0.74732251247800244354-0.17793644701386058601 i$~~~\\
15 & 0.048 &~~~$0.74734238325033472410-0.17792448779460261360 i$~~~\\
20 & 0.11 &~~~$0.74734333089004540317-0.17792456206654847739 i$~~~\\
25 & 0.17 &~~~$0.74734337404763909082-0.17792462415386593120 i$~~~\\
30 & 0.30 &~~~$0.74734337003564912217-0.17792463180189731963 i$~~~\\
\hline
\hline
\end{tabular}
\end{table*}

The numerical results for Regge-Wheeler effective potential Eq.~\eqref{eqRW} are presented in Tabs.~\ref{tab1}-\ref{tab3}.
By comparing the results obtained by the methods MV1 and its alternative MV2 in Tabs.~\ref{tab1} and~\ref{tab2}, it is observed that the accuracy is satisfactory and similar for the two approaches.
In particular, even at $N=5$, the numerical results are qualitatively reasonable.
Both approaches attain nine significant figures at the order $N=25$ by comparing against the results obtained by the continued fraction method.
The main difference between MV1 and MV2 resides in efficiency.
At $N=25$, a rational Chebyshev grid leads to an advantage in computational time by almost an order of magnitude.
As a reference, we also show in Tab.~\ref{tab3} the results of the original matrix method.
It is observed that the original matrix method is featured by its efficiency and reasonable accuracy but is unfortunately plagued by Runge's phenomenon at higher order~\cite{agr-qnm-lq-matrix-10}.
It is indicated that the computational time of the original approach MVO and that of MV2 is of the same order of magnitude.
Nevertheless, as discussed in~\cite{agr-qnm-lq-matrix-08}, at more significant orders, MVO does not guarantee better precision and deviates further from the correct value.

The scalar perturbations in asymptotically anti-de Sitter spacetime are carried out with the parameters $s=0$, $r_h=\frac{4}{5}$, $\ell = 0$, and $L=1$, where the horizon radius $r_+$ is related to the parameter $r_0$ by the relation $r_0=r_++r_+^3$.
The master equation Eq.~\eqref{qnmeq} in the ingoing Eddington coordinates yields~\cite{agr-qnm-HH-01}
\bqn\label{eqEddr}
F(r)\frac{d^2}{dr^2}\Phi(r)+\left[F'(r) - 2i\omega\right]\frac{d}{dr}\Phi(r) -\tilde{V}_{\mathrm{eff}}(r)\Phi(r)=0 ~,
\eqn
where
$\tilde{V}_{\mathrm{eff}}(r) =V_{\mathrm{eff}}/F(r)$.
The latter further simplies to
\bqn\label{eqRWr2}
\begin{aligned}
&\left[r^2(r+r^3-r_+(1+r_+^2))\right]\Phi''(r)+\left[r(2r^3+r_++r_+^3-2ir^2\omega)\right]\Phi'(r)\\
&+\left[-\ell(\ell+1) r-2r^3-r_+(1+r_+^2)\right]\Phi(r)=0 .
\end{aligned}
\eqn

Again, by introducing the variables $x$
\bqn
x = \frac{r_+}{r} ,
\eqn
which takes ``$1$'' and ``$0$'' at the horizon and spatial infinity, and
\bqn
y = 2\frac{r_+}{r} - 1 ,
\eqn
which manifestly attains ``$\pm 1$'' at the relevant boundaries, the master equation Eq.~\eqref{eqRWr2} gives
\bqn\label{eqRWx2}
\begin{aligned}
&\left[(x-1) x^4+r_+^2 x^2(x^3-1)\right] R''(x)+\left[x^2(-2x+3(1+r_+^2)x^2-2ir_+\omega)\right]R'(x)\\
&+\left[x^2(\ell(\ell+1)+x) + r_+^2(2+x^3)\right]R(x)=0 
\end{aligned}
\eqn
in $x$, where $R(x)=\Phi(r(x))$ for the standard matrix method and 
\bqn\label{eqRWy2}
\begin{aligned}
&\left[(y-1)(y+1)^2((y+1)^2+r_+^2(7+4y+y^2))\right]P''(y)+\left[(y+1)^2(-1+2y+3y^2+3r_+^2(1+y)^2i-8i r_+\omega\right]P'(y)\\
&+\left[2\ell(\ell+1)(y+1)^2+(y+1)^3+r_+ (17+3y+3y^2+y^3)\right]P(y)=0 .
\end{aligned}
\eqn
in variable $y$ for Chebyshev grid, where $P(y)=\Phi(r(y))$.

As a reference for the accurate value regarding the calculations presented below, the quasinormal frequency for the scalar QNMs in asymptotically anti-de Sitter spacetime is also evaluated by employing the HH method~\cite{agr-qnm-HH-01} at $100$th order.
The value is found to be 
\bq
\omega_\mathrm{HH} = 2.5877873521223829073-2.1304244788049905744 i .\nb
\eq

\begin{table*}[ht]
\caption{\label{tab4} The calculated fundamental mode employing the matrix method on the Chebyshev grid (MV1) for the scalar perturbations in anti-de Sitter spacetime by using different grid numbers.}
\centering
\begin{tabular}{c c c}
         \hline\hline
\text{Grid number }$N$ & \text{Time (seconds)} & $\omega$  \\
\hline
\hline
5  & 0.026 &~~~$2.0589166183236020389-1.8731979777909703338 i$~~~\\
10 & 0.25 &~~~$2.6217938827657266554-2.1392534574148980456 i$~~~\\
15 & 0.74 &~~~$2.5877996957117039078-2.1302724971621405316 i$~~~\\
20 & 2.17 &~~~$2.5877868571231347524-2.1304243983739371818 i$~~~\\
25 & 3.72 &~~~$2.5877873515744241392-2.1304244800617923928 i$~~~\\
\hline
\hline
\end{tabular}
\end{table*}

\begin{table*}[ht]
\caption{\label{tab5} The same as Tab.~\ref{tab4} but employing the matrix method implemented on the {\it rational} Chebyshev grid (MV2) as described in the text.}
\centering
\begin{tabular}{c c c}
         \hline\hline
\text{Grid number }$N$ & \text{Time (seconds)} & $\omega$  \\
\hline
\hline
5  & 0.012 &~~~$2.0586702116981006769-1.8734873515222502160 i$~~~\\
10 & 0.029 &~~~$2.6221822675836151997-2.1380918138026802981 i$~~~\\
15 & 0.042 &~~~$2.5878081024973376445-2.1302742499421558473 i$~~~\\
20 & 0.096 &~~~$2.5877868659317479198-2.1304243955697301531 i$~~~\\
25 & 0.14  &~~~$2.5877873514963902077-2.1304244801628264561 i$~~~\\
\hline
\hline
\end{tabular}
\end{table*}

\begin{table*}[ht]
\caption{\label{tab6} The calculated fundamental mode employing the original matrix method on the uniform grid (MVO) for the scalar perturbations in anti-de Sitter spacetime by using different grid numbers.}
\centering
\begin{tabular}{c c c}
         \hline\hline
\text{Grid number }$N$ & \text{Time (seconds)} & $\omega$  \\
\hline
\hline
5  & 0.012 &~~~$2.2599803837541830893-1.5997852644997328867 i$~~~\\
10 & 0.022 &~~~$2.4263308174531993102-2.1544879778417491785 i$~~~\\
15 & 0.046 &~~~$2.5888784011371891527-2.1369475594335492285 i$~~~\\
20 & 0.079 &~~~$2.5879260875538251257-2.1304816765435536055 i$~~~\\
25 & 0.14 &~~~$2.5877898461365884221-2.1304230185934063323 i$~~~\\
30 & 0.21 &~~~$2.5877873607213837181-2.1304244297555327022 i$~~~\\
\hline
\hline
\end{tabular}
\end{table*}

The numerical results for the QNMs of the scalar perturbations in asymptotically anti-de Sitter spacetime are presented in Tabs.~\ref{tab4}-\ref{tab6}.
The results obtained by the methods MV1 and its alternative MV2 shown in Tabs.~\ref{tab4} and~\ref{tab5} indicate that both approaches furnish similarly satisfactory precision.
In particular, for $N\ge 15$, reasonably accurate results are obtained.
Both approaches attain nine significant figures at the order $N=25$ by comparing against the results obtained by the HH method.
The main difference between MV1 and MV2 resides in the computational time.
At $N=25$, by employing the rational Chebyshev grid, MV2 leads to an advantage in computational time by more than an order of magnitude.
Again, as a reference, we also show in Tab.~\ref{tab6} the results obtained by the original matrix method.
It is observed that the original matrix method furnishes reasonable efficiency and accuracy.
It is noted that the computational time of the original approach MVO and MV2 is of the same order of magnitude.
For more significant $N$, MVO starts to deviate from the correct value and no longer guarantees precision.
As discussed above, the latter is attributed to Runge's phenomenon that occurs at higher order~\cite{agr-qnm-lq-matrix-10}.

The above results indicate a trade-off between efficiency and accuracy.
The original matrix method MVO is efficient, but Runge's phenomenon at higher orders potentially plagues its precision. 
Nonetheless, it is reasonably robust and a good candidate for most applications with a moderate grid number.
The matrix method on rigorous Chebyshev grid MV1 is somewhat constrained by the computational time.
The matrix method implemented on rational Chebyshev grid MV2 seems to possess both desirable features and therefore is more accessible from a practical viewpoint.

\section{Concluding remarks} \label{section5}

Runge's phenomenon poses a challenge to the accuracy of the matrix method at higher orders, which has been implemented on a uniform grid in most of its applications. 
In the present work, we explore this issue and generalize the matrix method to adopt one of the best-known remedies to Runge's phenomenon, namely, the Chebyshev grid.
As a nonuniform grid, adapting to the Chebyshev grid guarantees an effective suppression of Runge's phenomenon, leading to faster convergence.
However, it is found that while such an implementation is favorable from a mathematical point of view, the increase in precision does not necessarily compensate for the penalty in computational time.
In practice, though subject to divergence at higher order, the original matrix method is reasonably robust and serves most applications when a moderate grid number is assumed.
In this regard, there is a trade-off between these two aspects: computational time and accuracy.
We argue that an alternative version of the matrix method, which takes nodes with rational values while being as close to the Chebyshev grid as possible, furnishes a desirable outcome.
Numerical studies have been carried out for the axial gravitational and scalar perturbations in asymptotically flat and anti-de Sitter spacetimes.
The conclusion drawn by the numerical calculations has also been validated by considering different types of perturbations, angular momenta, among others.

\section*{Acknowledgments}
We acknowledge insightful discussions with Kai Lin.
This work is supported by the National Natural Science Foundation of China (NNSFC) under contract Nos. 11805166, 11925503, and 12175076.
We also gratefully acknowledge the financial support from
Funda\c{c}\~ao de Amparo \`a Pesquisa do Estado de S\~ao Paulo (FAPESP),
Funda\c{c}\~ao de Amparo \`a Pesquisa do Estado do Rio de Janeiro (FAPERJ),
Conselho Nacional de Desenvolvimento Cient\'{\i}fico e Tecnol\'ogico (CNPq),
Coordena\c{c}\~ao de Aperfei\c{c}oamento de Pessoal de N\'ivel Superior (CAPES),
A part of this work was developed under the project Institutos Nacionais de Ci\^{e}ncias e Tecnologia - F\'isica Nuclear e Aplica\c{c}\~{o}es (INCT/FNA) Proc. No. 464898/2014-5.
This research is also supported by the Center for Scientific Computing (NCC/GridUNESP) of S\~ao Paulo State University (UNESP).

\bibliographystyle{h-physrev}
\bibliography{references_qian, references_mm_citations}

\end{document}